\documentclass[aps,prl,twocolumn,superscriptaddress,showpacs]{revtex4-1}
\usepackage{amsmath,amssymb}
\usepackage{bm}
\usepackage[dvipdfm]{graphicx}

\usepackage{color}

\def\oH{\hat{H}}
\def\oHS{\hat{H}_{\text{S}}}
\def\oHR{\hat{H}_{\text{R}}}
\def\oHSR{\hat{H}_{\text{SR}}}

\def\oa{\hat{a}}
\def\oad{\hat{a}^{\dagger}}
\def\oc{\hat{c}}
\def\ocd{\hat{c}^{\dagger}}
\def\ob{\hat{b}}
\def\obd{\hat{b}^{\dagger}}
\def\oPsi{\hat{\varPsi}} 
\def\oPsid{\hat{\varPsi}^{\dagger}}

\def\Hc{\mathrm{H.c.}}

\def\ii{\mathrm{i}}
\def\dd{\mathrm{d}}

\def\Ek{E_{\bm{k}}}
\def\uk{u_{\bm{k}}}
\def\vk{v_{\bm{k}}}
\def\Dk{\varDelta_{\bm{k}}}

\def\U#1{U'_{\bm{#1}}}
\def\kB{k_{\text{B}}}
\def\muB{\mu_{\text{B}}}

\def\fehB{f_{\text{e/h}}^{\text{B}}}

\def\mueB{\mu_{\text{e}}^{\text{B}}}
\def\muhB{\mu_{\text{h}}^{\text{B}}}
\def\muehB{\mu_{\text{e/h}}^{\text{B}}}

\def\xiehk#1{\tilde{\xi}_{\text{eh},\bm{k}}^{#1}}

\def\Eg{E_{\text{g}}}

\begin{document}
\title{Second Thresholds in BEC-BCS-Laser Crossover of Exciton-Polariton Systems}

\author{Makoto Yamaguchi}
\altaffiliation{E-mail: yamaguchi@acty.phys.sci.osaka-u.ac.jp}
\affiliation{Department of Physics, Osaka University, 1-1 Machikaneyama, Toyonaka, Osaka 560-0043, Japan}
\author{Kenji Kamide}
\affiliation{Department of Physics, Osaka University, 1-1 Machikaneyama, Toyonaka, Osaka 560-0043, Japan}
\author{Ryota Nii}
\affiliation{Department of Physics, Osaka University, 1-1 Machikaneyama, Toyonaka, Osaka 560-0043, Japan}
\author{Tetsuo Ogawa}
\affiliation{Department of Physics, Osaka University, 1-1 Machikaneyama, Toyonaka, Osaka 560-0043, Japan}
\affiliation{Photon Pioneers Center, Osaka University, 2-1 Yamada-oka, Suita, Osaka 565-0871, Japan}
\author{Yoshihisa Yamamoto}
\affiliation{National Institute of Informatics, 2-1-2 Hitotsubashi, Chiyoda-ku, Tokyo 101-8403, Japan}
\affiliation{E. L. Ginzton Laboratory, Stanford University, Stanford, California 94305, USA}

\date{\today}

\begin{abstract}
The mechanism of second thresholds observed in several experiments is theoretically revealed by studying the BEC-BCS-laser crossover in exciton-polariton systems.
We found that there are two different types for the second thresholds; one is a crossover within quasi-equilibrium phases and the other is into non-equilibrium (lasing).
In both cases, the light-induced band renormalization causes gaps in the conduction and valence bands, which indicates the existence of bound electron-hole pairs in contrast to earlier expectations.
We also show that these two types can be distinguished by the gain spectra.
\end{abstract}


\maketitle
In semiconductor exciton-polariton systems, Bose-Einstein condensation (BEC) of exciton-polaritons has been observed in recent years~\cite{Deng02, Kasprzak06, Balili07, Utsunomiya08, Horikiri10,Belykh13}.
A hot issue is, now, how the exciton-polariton BEC, a thermal equilibrium phenomenon, changes into the lasing operation resulting from the electron-hole (e-h) plasma gain~\cite{Chow99}, which is essentially a non-equilibrium phenomenon~\cite{Bajoni08, Kasprzak08}.
Earlier experiments show that there are two distinct thresholds when increasing the excitation density: the first one is the critical density for the BEC~\cite{Balili09, Nelsen09} and the second one is recognized as the standard lasing~\cite{Dang98, Balili09, Nelsen09, Tempel12, Tsotsis12, Kamman12}.
In most cases, the second-threshold mechanism is explained by a shift into the weak coupling regime due to dissociations of Coulomb-bound e-h pairs (excitons) into the \mbox{e-h} plasma.
However, there is no convincing discussion why such dissociations lead to non-equilibration of the system essential for lasing.
As another possibility, a new ordered state involving Bardeen-Cooper-Schrieffer (BCS) -like correlation is also speculated~\cite{Deng10}.
The second threshold is, thus, currently subject to intense debate.
In this letter, our purpose is to reveal the mechanism of the second threshold by studying the BEC-BCS-laser crossover theories~\cite{Szymanska06, Keeling10, Yamaguchi12}.
As a result, we found that there are two different types for the second threshold; one is a crossover into photonic polariton BEC (quasi-equilibrium)~\cite{Kamide10, Byrnes10} and the other is into lasing (non-equilibrium)~\cite{Note1}.
In both cases, the light-induced band renormalization causes gaps inside the conduction and valence bands, which indicates that there are still light-induced e-h pairs even after the second thresholds, in contrast to the above scenario.
We also show that these two types can be distinguished by the gain spectra.

Our model Hamiltonian is $\oH = \oHS + \oHR + \oHSR$, where
\begin{align}%%%%%%%%%%%%%%%%%%%%%%%%%%%%%%%%%%%%%%% Equations %%%%%%%%%%%%%%%%%%%%%%%%%%%%%%%%%%%%%%%%%%%%%%%%
\oHS & = \hbar\sum_{\alpha,\bm{k}}\xi_{\alpha,\bm{k}}\ocd_{\alpha,\bm{k}}\oc_{\alpha,\bm{k}}
       + \hbar\sum_{\bm{q}}\xi_{\text{ph},\bm{q}}\oad_{\bm{q}}\oa_{\bm{q}} \displaybreak[0] \nonumber \\ 
     & + \frac{1}{2}\sum_{\bm{k},\bm{k}',\bm{q}}\sum_{\alpha,\alpha'}\U{q}\ocd_{\alpha,\bm{k}+\bm{q}}
		\ocd_{\alpha',\bm{k}'-\bm{q}}\oc_{\alpha',\bm{k}'}\oc_{\alpha,\bm{k}} \displaybreak[0] \nonumber \\ 
     & -\hbar\sum_{\bm{k},\bm{q}}(g^*\oa_{\bm{q}}\ocd_{\text{c},\bm{k}+\bm{q}}\oc_{\text{v},\bm{k}}+\Hc),
\label{eq:oHS} \\
\oHR & = \hbar\sum_{\alpha,\bm{k}}\xi_{\alpha,\bm{k}}^{\text{B}}\obd_{\alpha,\bm{k}}\ob_{\alpha,\bm{k}}
       + \hbar\sum_{\bm{p}}\xi_{\bm{p}}^{\text{B}}\oPsid_{\bm{p}}\oPsi_{\bm{p}},
\label{eq:oHR} \\
\oHSR & = \hbar\sum_{\alpha,\bm{k},\bm{q}}\varGamma_{\bm{k}}^{\alpha}\ocd_{\alpha,\bm{k}}\ob_{\alpha,\bm{q}}
       + \hbar\sum_{\bm{p},\bm{q}}\zeta_{\bm{q}}\oad_{\bm{q}}\oPsi_{\bm{p}}+\Hc,
\label{eq:oHSR}
\end{align}%%%%%%%%%%%%%%%%%%%%%%%%%%%%%%%%%%%%%%%%%%%%%%%%%%%%%%%%%%%%%%%%%%%%%%%%%%%%%%%%%%%%%%%%%%%%%%%%%%%%
are the system, reservoir, and their interaction Hamiltonians, respectively, where $\alpha, \alpha' \in \{\text{c, v}\}$~\cite{Yamaguchi12}.
In Eq.~\eqref{eq:oHS}, $\oc_{\text{c},\bm{k}}$ ($\oc_{\text{v},\bm{k}}$) is the conduction (valence) band electron annihilation operators with the electronic dispersion $\xi_{\text{c/v},\bm{k}} \equiv \omega_{\text{c/v},\bm{k}} \mp \hbar^{-1}\mu/2$, $\oa_{\bm{q}}$ is the cavity photon annihilation operator with the photonic dispersion $\xi_{\text{ph},\bm{q}} \equiv \omega_{\text{ph},\bm{q}} - \hbar^{-1}\mu$ and $\hbar^{-1}\mu$ is an oscillation frequency of the photon and polarization fields, which corresponds to the energy of the main peak in photoluminescence.
The carriers are interacting with each other through the Coulomb interaction $U'_{\bm{q}}$, and they can emit photons via the light-matter coupling constant $g$.
Similarly, in Eqs.~\eqref{eq:oHR} and \eqref{eq:oHSR}, $\ob_{\text{c},\bm{k}}$ and $\ob_{\text{v},\bm{k}}$ denote fermion annihilation operators of pumping baths and $\oPsi_{\bm{p}}$ is a boson annihilation operator of free-space vacuum fields.
In this model, Eqs.~\eqref{eq:oHR} and \eqref{eq:oHSR} are responsible for the incoherent fermionic pumping and photon decay \cite{suppl}.
Based on the above Hamiltonians, we focus on steady states described by the polarization function $p_{\bm{k}} \equiv \langle \ocd_{\text{v},\bm{k}}\oc_{\text{c},\bm{k}}\rangle$, the number of electrons $n_{\text{e},\bm{k}} \equiv \langle\ocd_{\text{c},\bm{k}}\oc_{\text{c},\bm{k}}\rangle$ and holes $n_{\text{h},-\bm{k}} \equiv 1 - \langle\ocd_{\text{v},\bm{k}}\oc_{\text{v},\bm{k}}\rangle$, and a coherent photon field formed in the $\bm{q} = 0$ state $\langle\oa_{\bm{q}}\rangle = \delta_{\bm{q},0}a_0$ with the oscillation frequency $\hbar^{-1}\mu$. 
These are the minimum variables for describing the BEC, BCS states, and semiconductor lasers.

The band renormalization of the e-h system can, then, be conveniently studied by the poles of the single-particle spectral function $A_{\alpha \alpha}(\nu;\bm{k})$.
Within the Hartree-Fock approximation (HFA), a standard Green's function technique yields
\begin{align}%%%%%%%%%%%%%%%%%%%%%%%%%%%%%%%%%%%%%%% Equations %%%%%%%%%%%%%%%%%%%%%%%%%%%%%%%%%%%%%%%%%%%%%%%%
A_{\text{cc}/\text{vv}}(\nu;\bm{k}) & = 2|\uk|^2 L(\nu,\mp \Ek) + 2|\vk|^2 L(\nu,\pm \Ek),
\label{eq:Accvv}%%%%%%%%%%%%%%%%%%%%%%%%%%%%%%%%%%%%%%%%%%%%%%%%%%%%%%%%%%%%%%%%%%%%%%%%%%%%%%%%%%%%%%%%%%%%%%%
\end{align}
where $\sqrt{2}\uk \equiv [1 + \xiehk{+}/\Ek]^{1/2}$ and $\sqrt{2}\vk \equiv e^{\ii\theta_{\bm{k}}} [1 - \xiehk{+}/\Ek]^{1/2}$ with $\Ek \equiv [(\xiehk{+})^2 + |\Dk|^2]^{1/2}$ are the Bogoliubov coefficients, $L(\nu,\pm \Ek) \equiv \hbar\gamma/[(\hbar\nu-\hbar\xiehk{-} \pm \hbar\Ek)^2+(\hbar\gamma)^2]$ is the Lorentz function, $\Dk = |\Dk|e^{\ii\theta_{\bm{k}}} \equiv g^*a_0 + \hbar^{-1}\sum_{\bm{k}'}U'_{\bm{k}'-\bm{k}}p_{\bm{k}'}$ is a composite order parameter, and $\gamma$ is the thermalization rate of the e-h system \cite{suppl}.
In the derivation, the notation is transformed into the e-h picture with $\xiehk{\pm} \equiv (\tilde{\xi}_{\text{e},\bm{k}} \pm \tilde{\xi}_{\text{h},\bm{k}})/2$ where $\tilde{\xi}_{\text{e/h},\bm{k}} \equiv \omega_{\text{e/h},\bm{k}} + \varSigma_{\text{e/h},\bm{k}}^{\text{BGR}} - \hbar^{-1}\mu/2$ describes the single particle energy renormalized by the Coulomb interactions $\hbar\varSigma_{\text{e/h},\bm{k}}^{\text{BGR}} \equiv - \sum_{\bm{k}'}U'_{\bm{k}'-\bm{k}}n_{\text{e/h},\bm{k}'}$.

\begin{figure}[tbp]%%%%%%%%%%%%%%%%%%%%%%%%%%%%%%%%%%%%%%% Figure 1 %%%%%%%%%%%%%%%%%%%%%%%%%%%%%%%%%%%%%%%%%%%%%%%%
\begin{center}
\includegraphics{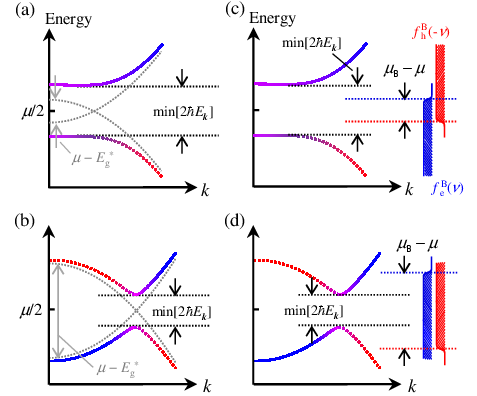}
\vspace{-0.6cm}
\end{center}
\caption{Renormalized conduction bands for (a) large ($|\mu-E_{\text{g}}^*| \lesssim \min[2\hbar\Ek]$) and (b) small ($\min[2\hbar\Ek] \lesssim \mu-E_{\text{g}}^*$) gap energies.
Here, $E_{\text{g}}^* \equiv E_{\text{g}} + \hbar\varSigma_{\text{e},\bm{k}=0}^{\text{BGR}} + \hbar\varSigma_{\text{h},\bm{k}=0}^{\text{BGR}}$ and the gray dashed lines show the energies by ignoring $\Dk$ in $\Ek$.
Relations to $f^{\text{B}}_{\text{e/h}}(\pm\nu)$  are also illustrated in panel (c) for quasi-equilibrium (condition (I)) and panel (d) for lasing (condition (II)).
$\omega_{\text{e},k}=\omega_{\text{h},k}$ and $\mueB = \muhB$ are assumed.
}
\label{fig:fig1}
\vspace{-0.5cm}
\end{figure}%%%%%%%%%%%%%%%%%%%%%%%%%%%%%%%%%%%%%%%%%%%%%%%%%%%%%%%%%%%%%%%%%%%%%%%%%%%%%%%%%%%%%%%%%%%%%%%%%%%%%%

In Eq.~\eqref{eq:Accvv}, there are remarkable similarities to superconductivities \cite{Abrikosov75}.
It is then clear that $\min[2\hbar\Ek]$ represents the gap energy opened at $\mu/2$ in the renormalized conduction and valence bands (typically Fig.~\ref{fig:fig1} (a) and (b)).
Such a picture is well-known for e.g. the BEC-BCS crossover but, now, one should notice that Eq.~\eqref{eq:Accvv} is also applicable for {\it lasing}~\cite{Note1} because thermal equilibrium is not required in the derivation.
In this case, $\mu$ is not the chemical potential but the laser frequency.
Furthermore, the gap is opened at $\mu/2$ whenever lasing because $a_0 \neq 0$ and $p_{\bm{k}} \neq 0$ result in $\min[2\hbar\Ek] \neq 0$.
The origin of the gap is analogous to the Rabi splitting in resonance fluorescence \cite{Scully77, Schmitt-Rink88, Henneberger92, Horikiri}.
This can be understood from the expression of $\min[2\hbar\Ek] = 2\hbar |g|\sqrt{n_{\text{ph}}}$ ($n_{\text{ph}} \equiv |a_0|^2$) obtained by assuming free electrons ($U'_{\bm{q}} = 0$) with $\mu > E_{\text{g}}$ ($E_{\text{g}}$; the bare band gap energy), which is equivalent to the Rabi frequency in resonance fluorescence.
Hence, it is worth noting that the existence of the gap indicates that light-induced e-h pairs do exist whenever lasing even though there is no e-h pair before lasing.
This is one of our important results despite the quite simple analysis.

For later convenience, two typical situations for large and small gap energies are shown in Fig.~\ref{fig:fig1} (a) and (b), respectively.
In Fig.~\ref{fig:fig1} (a), the renormalized conduction band has a gap around $k \approx 0$ with flattened dispersions because e-h band mixing occurs for large $k$-regions.
In contrast, in Fig.~\ref{fig:fig1} (b), the renormalization is mainly focused on particular $k$-regions.
In both cases, the renormalized bands have gaps at $\mu/2$ and the same holds for the valence band (not shown).

In order to discuss the second-threshold mechanism, however, the unknown variables $a_0$, $p_{\bm{k}}$, $n_{\text{e},\bm{k}}$, $n_{\text{h},\bm{k}}$, and $\mu$ in Eq.~\eqref{eq:Accvv} should be determined in a comprehensive way including BEC, BCS and laser physics \cite{Szymanska06, Keeling10, Yamaguchi12}. Within the HFA, the simultaneous steady-state equations, derived from Eqs.~\eqref{eq:oHS}-\eqref{eq:oHSR}, can formally be written as
\begin{align}%%%%%%%%%%%%%%%%%%%%%%%%%%%%%%%%%%%%%%% Equations %%%%%%%%%%%%%%%%%%%%%%%%%%%%%%%%%%%%%%%%%%%%%%%%
&\partial_ta_0=0=-\ii\xi_{\text{ph},0}a_0+\ii g \textstyle{\sum_{\bm{k}}} p_{\bm{k}}-\kappa a_0,
\label{eq:a0}\\
&\partial_tp_{\bm{k}}=0=-2\ii\xiehk{+}p_{\bm{k}}-\ii\Dk N_{\bm{k}}-2\gamma (p_{\bm{k}}-p_{\bm{k}}^0),
\label{eq:pk}\\
&\partial_tn_{\text{e/h},\bm{k}}=0=-2\Im[\Dk p_{\bm{k}}^{*}]-2\gamma(n_{\text{e/h},\bm{k}}-n_{\text{e/h},\bm{k}}^0),
\label{eq:nehk}%%%%%%%%%%%%%%%%%%%%%%%%%%%%%%%%%%%%%%%%%%%%%%%%%%%%%%%%%%%%%%%%%%%%%%%%%%%%%%%%%%%%%%%%%%%%%%%
\end{align}
where $N_{\bm{k}} \equiv n_{\text{e},\bm{k}}+n_{\text{h},\bm{k}}-1$ is the population inversion and $\kappa$ is the photon loss rate.
Note that Eqs.~\eqref{eq:a0}-\eqref{eq:nehk} have well-known forms of the Maxwell-Semiconductor-Bloch equations (MSBE) under the relaxation time approximation (RTA) if $n_{\text{e/h},\bm{k}}^0$ is replaced by the Fermi distribution with $p_{\bm{k}}^0=0$ \cite{Chow99,Henneberger92,Yamaguchi12,Chow02}.
In general, the MSBE under the RTA can describe the physics of semiconductor lasers but cannot describe the BEC and BCS states.
However, the key point here is that Eqs.~\eqref{eq:a0}-\eqref{eq:nehk} become able to describe the BEC, BCS, and laser physics in a unified way when $p_{\bm{k}}^0$ and $n_{\text{e/h},\bm{k}}^0$ are described by
\begin{align*}%%%%%%%%%%%%%%%%%%%%%%%%%%%%%%%%%%%%%%% Equations %%%%%%%%%%%%%%%%%%%%%%%%%%%%%%%%%%%%%%%%%%%%%%%%
&p_{\bm{k}}^0 \equiv \ii\textstyle\int\frac{\dd[\hbar\nu]}{2\pi}(G_{\text{cv},\bm{k}}^{\text{R}}(\nu )( 1-f^{\text{B}}_{\text{h}}(-\nu ))-G_{\text{vc},\bm{k}}^{\text{R*}}(\nu )f^{\text{B}}_{\text{e}}(\nu )),\\
&n_{\text{e/h},\bm{k}}^0 \equiv \textstyle\int\frac{\dd[\hbar\nu]}{2\pi}\fehB(\nu) A_{\text{cc/vv}}(\pm\nu;\bm{k}),
%%%%%%%%%%%%%%%%%%%%%%%%%%%%%%%%%%%%%%%%%%%%%%%%%%%%%%%%%%%%%%%%%%%%%%%%%%%%%%%%%%%%%%%%%%%%%%%
\end{align*}
where $\fehB(\nu) \equiv [\exp(\beta(\hbar\nu - (\muehB-\mu/2))) + 1]^{-1}$ is the bath Fermi distribution with the chemical potential $\muehB$.
The exact expression of $G_{\alpha\alpha',\bm{k}}^{\text{R}}$ is given in the Supplemental Material~\cite{suppl}.
Then, by assuming $\omega_{\text{e},\bm{k}} = \omega_{\text{h},\bm{k}}$ and a charge neutrality $\mueB = \muhB$ with $\muB \equiv \mueB + \muhB$, it can be shown that Eqs.~\eqref{eq:a0}-\eqref{eq:nehk} can recover the BCS gap equation when (I) $\min[2\hbar\Ek] \gtrsim \muB - \mu + 2\hbar\gamma + 2\kB T$ (quasi-equilibrium).
In contrast, there appear $k$-regions described by the MSBE when (II) $\muB-\mu \gtrsim \min[2\hbar\Ek] + 2\hbar\gamma + 2\kB T$ (lasing; non-equilibrium).
The physical meanings of these conditions are discussed in detail in Ref.~\cite{Yamaguchi12} (see also the Supplemental Material~\cite{suppl}) and not repeated here.
Instead, these conditions are illustrated in Fig.~\ref{fig:fig1} (c) and (d) in relation to $\fehB(\pm\nu)$ for quasi-equilibrium and lasing conditions, respectively.
It is, then, clear that the system enters into lasing phases when $\muB - \mu$ roughly goes beyond the energy gap $\min[2\hbar\Ek]$ by ignoring the broadening due to $\gamma$ and $T$.
\begin{figure}[tbp]%%%%%%%%%%%%%%%%%%%%%%%%%%%%%%%%%%%%%%% Figure 2 %%%%%%%%%%%%%%%%%%%%%%%%%%%%%%%%%%%%%%%%%%%%%%%%
\begin{center}
\includegraphics{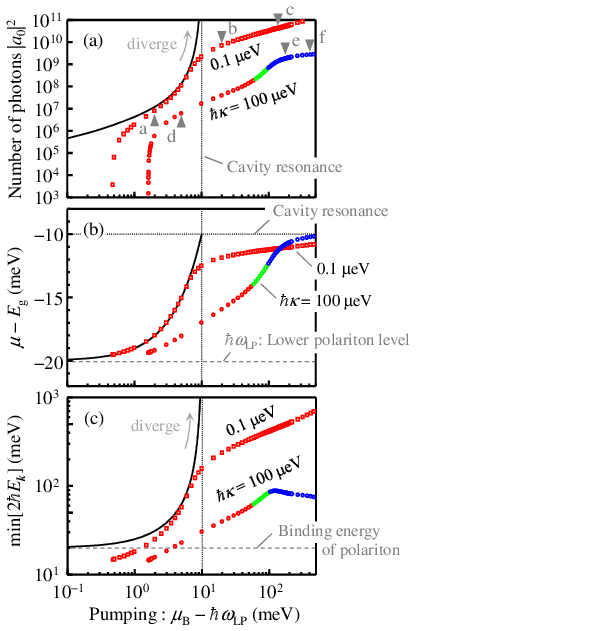}
\vspace{-0.6cm}
\end{center}
\caption{Numerical solutions of (a) the number of photons $|a_0|^2$, (b) the oscillation frequency $\mu$, and (c) the gap energy $\min[2\hbar\Ek]$ as a function $\muB$ for $\hbar\kappa = 0.1~{\rm\mu eV}$ and $100~{\rm\mu eV}$.
Red and blue colors are used when satisfying the quasi-equilibrium condition (I) and the lasing condition (II) in the text, respectively. Green colors are used when neither of them are satisfied. For comparison, black solid lines show the results by thermal-equilibrium theories \cite{Kamide10, Byrnes10}.}
\label{fig:fig2}
\vspace{-0.6cm}
\end{figure}%%%%%%%%%%%%%%%%%%%%%%%%%%%%%%%%%%%%%%%%%%%%%%%%%%%%%%%%%%%%%%%%%%%%%%%%%%%%%%%%%%%%%%%%%%%%%%%%%%%%%%

Based on the above formalism, we have performed numerical calculations, where the cavity level (= $\hbar\omega_{\text{ph},0}$) is in resonance with the (1S) exciton level located at 10 meV below $E_{\text{g}}$ and the lower polariton level $\hbar\omega_{\text{LP}}$ is formed at 20 meV below $E_{\text{g}}$~\cite{suppl}.
For $\hbar\kappa$, we have used values of $0.1~{\rm\mu eV}$ and $100~{\rm\mu eV}$ to study the effects of non-equilibrium.
We note, however, that $\hbar\kappa = 100~{\rm\mu eV}$ is a reasonable value in current experiments.
Fig.~\ref{fig:fig2} shows the calculated results of $|a_0|^2$, $\mu$, and $\min[2\hbar\Ek]$ as a function of $\muB$, the pumping parameter.
In the case of the equilibrium theories, $|a_0|^2$ diverges in the limit of $\muB \to \hbar\omega_{\text{ph},0}$ because it is preferable to increase photons rather than electrons and holes due to the phase space filling effects.
As a result, the photonic polariton BEC is achieved by the photon-mediated e-h attraction \cite{Kamide10, Byrnes10}.
In contrast, in the case with finite pumping and losses (plots), the behaviors are different in many aspects.
Focusing on the plots for $\hbar\kappa = 0.1~{\rm\mu eV}$, two distinct thresholds can be seen ($\muB-\hbar\omega_{\text{LP}} \approx 5.0\times10^{-1}$  meV and $6.0\times10^{0}$ meV) in Fig.~\ref{fig:fig2} (a). 
At the same time, $\mu$ is gradually blue-shifted from $\hbar\omega_{\text{LP}}$ and then approaches the bare cavity resonance (Fig.~\ref{fig:fig2} (b)).
Similar qualitative behaviors also can be seen for $\hbar\kappa = 100~{\rm\mu eV}$.
These behaviors are consistent with experiments.
However, there is a crucial difference between the two; according to the above-mentioned conditions (I) and (II), all plots are in quasi-equilibrium for $\hbar\kappa = 0.1~{\rm\mu eV}$ but there are plots (blue) in lasing for $\hbar\kappa = 100~{\rm\mu eV}$ after the second threshold.
\begin{figure}[tbp]%%%%%%%%%%%%%%%%%%%%%%%%%%%%%%%%%%%%%%% Figure 3 %%%%%%%%%%%%%%%%%%%%%%%%%%%%%%%%%%%%%%%%%%%%%%%%
\begin{center}
\includegraphics{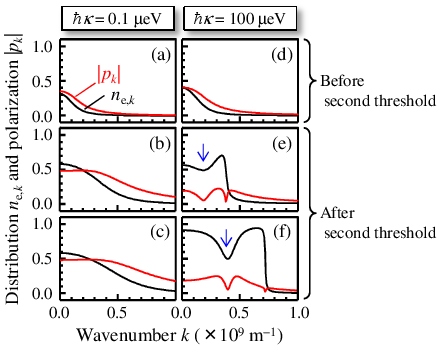}
\vspace{-0.6cm}
\end{center}
\caption{Calculated distributions $n_{\text{e},k}$ and polarizations $|p_k|$ corresponding to the triangles indicated by a-f in Fig. \ref{fig:fig2} (a). Panel (a)-(c) are for $\hbar\kappa = 0.1~{\rm\mu eV}$ and panel (d)-(f) are for $\hbar\kappa = 100~{\rm\mu eV}$.
Arrows show the kinetic hole burning.}
\label{fig:fig3}
\vspace{-0.6cm}
\end{figure}%%%%%%%%%%%%%%%%%%%%%%%%%%%%%%%%%%%%%%%%%%%%%%%%%%%%%%%%%%%%%%%%%%%%%%%%%%%%%%%%%%%%%%%%%%%%%%%%%%%%%%

The difference is also reflected in $n_{\text{e},k}$ and $p_k$, as shown in Fig.~\ref{fig:fig3}.
Before the second thresholds, $n_{\text{e},k}$ and $p_k$ for $\hbar\kappa = 0.1~{\rm\mu eV}$ (Fig.~\ref{fig:fig3} (a)) are similar to those for $\hbar\kappa = 100~{\rm\mu eV}$ (Fig.~\ref{fig:fig3} (d)).
However, after the second thresholds, $n_{\text{e},k}$ and $p_k$ are quite different, depending on the value of $\hbar\kappa$.
In the case of $\hbar\kappa = 0.1~{\rm\mu eV}$, $n_{\text{e},k}$ monotonically decreases as a function of $k$ and $p_k$ has a plateau $\approx$ 0.5 (Fig.~\ref{fig:fig3} (b) and (c)), which are the same features as the photonic polariton BEC in quasi-equilibrium \cite{Kamide10, Byrnes10}.
In contrast, in the case of $\hbar\kappa = 100~{\rm\mu eV}$, the kinetic hole burning appears as a signature of lasing and the Fermi surface is formed with the population inversion $n_{\text{e},k} > 0.5$ (Fig.~\ref{fig:fig3} (e)). 
For larger $\muB$, such behaviors become much more pronounced (Fig.~\ref{fig:fig3} (f)).
These results directly show that the second thresholds for $\hbar\kappa = 0.1~{\rm\mu eV}$ and $100~{\rm\mu eV}$ in Fig.~\ref{fig:fig2} (a) are formed by different mechanisms.

In fact, for $\hbar\kappa = 0.1~{\rm\mu eV}$, the second threshold is formed by the same mechanism as the photon divergence in the equilibrium theories, and therefore, it results from the crossover into the photonic polariton BEC.
In the present case, there are finite losses of cavity photons even if the system is in quasi-equilibrium.
As a result, the divergence is avoided and the second threshold appears instead.
After the second threshold, the monotonic increase of $\min[2\hbar\Ek]$ (Fig.~\ref{fig:fig2} (c)) indicates the enhancement of the light-induced e-h paring and expands the flattened region of dispersion in Fig.~\ref{fig:fig1} (a).
It is, then, clear that the plateau of $p_k \approx 0.5$ in Fig.~\ref{fig:fig3} (b) and (c) is formed by the e-h mixing around such flattened dispersions.
In the case of $\hbar\kappa = 100~{\rm\mu eV}$, on the other hand, the second threshold is related to the crossover into lasing, explained as follows.
Before the second threshold, the system stays in quasi-equilibrium (red circles in Fig.~\ref{fig:fig2}), where the relationship between the renormalized band and the pumping baths is well expressed in Fig.~\ref{fig:fig1} (c).
In this situation, the pumping is blocked inside the gap $\min[2\hbar\Ek]$.
However, by increasing the pumping $\muB$, $\muB-\mu$ exceed the gap, $\muB-\mu \gtrsim \min[2\hbar\Ek]$, and then, electrons above the gap can be supplied suddenly.
Such a feeding mechanism causes a rapid increase of photons, resulting in the second threshold.
Here, by ignoring the effects of $\gamma$ and $T$, this situation $\muB-\mu \gtrsim \min[2\hbar\Ek]$ is equivalent to the above-described condition (II) for the lasing phases.
Consequently, the second threshold is accompanied by the change into lasing (non-equilibrium).
By increasing the pumping further, $\mu$ is fixed around the cavity (Fig.~\ref{fig:fig2} (b)), $\min[2\hbar\Ek]$ is decreased (Fig.~\ref{fig:fig2} (c)), and the effective band gap $E_{\text{g}}^* \equiv E_{\text{g}} + \hbar\varSigma_{\text{e},\bm{k}=0}^{\text{BGR}} + \hbar\varSigma_{\text{h},\bm{k}=0}^{\text{BGR}}$ shrinks, of course.
The lasing situation is then well captured in Fig.~\ref{fig:fig1} (d), where the gap $\min[2\hbar\Ek]$ is decreased but still opened around the laser frequency.
The decrease of the gap for $\muB-\mu \gtrsim \min[2\hbar\Ek]$ implies that the particle flux $\muB-\mu$ beyond $\min[2\hbar\Ek]$ act toward e-h pair breaking but the e-h pairs cannot be fully dissociated because $\min[2\hbar\Ek] \ne 0$.
As a result, light-induced e-h pairs are still formed around the laser frequency, typically around the energy regions of the kinetic hole burning (Fig.~\ref{fig:fig3} (e) and (f)).
This is, in turn, somewhat analogous to the e-h Cooper pairs formed around the Fermi energy, i.e. weakly correlated e-h pairs in momentum space.
The difference is that the e-h pairs are formed around the laser frequency rather than the Fermi energy.

These results indicate that it would be reasonable to explain the second thresholds reported in current experiments by the crossover into lasing because $\kappa = 100~{\rm\mu eV}$ is a reasonable value for them, in agreement with earlier explanations \cite{Dang98, Balili09, Tempel12, Kamman12, Tsotsis12}.
Our theory, however, show that the crossover is not accompanied by the dissociations of bound e-h pairs.
Instead, the pairing mechanism changes into the light-induced one around the laser frequency.
This is in contrast to the commonly accepted ideas but a natural picture of lasing.

\begin{figure}[tbp]%%%%%%%%%%%%%%%%%%%%%%%%%%%%%%%%%%%%%%% Figure 4 %%%%%%%%%%%%%%%%%%%%%%%%%%%%%%%%%%%%%%%%%%%%%%%%
\begin{center}
\includegraphics{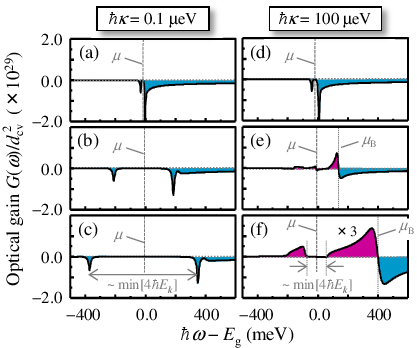}
\vspace{-0.6cm}
\end{center}
\caption{Optical gain spectra. Panel (a)-(f) corresponds to Fig.~\ref{fig:fig3} (a)-(f), respectively. Positive (red) and negative (blue) values represents the gain and absorption, respectively.}
\label{fig:fig4}
\vspace{-0.6cm}
\end{figure}

We have thus discussed the two different types of the second threshold.
However, it is difficult to directly distinguish them by the excitation dependence of the number of photons (Fig.~\ref{fig:fig2} (a)), in principle.
Therefore, we finally study the measurable optical gain spectra $G(\omega)$ \cite{Takahashi05, Yoshita12} by assuming an additional perturbative Hamiltonian $\hat{H}'(t) = -F(t)\sum_{\bm{k}}d_{\text{cv}}(\ocd_{\text{c},\bm{k}}\oc_{\text{v},\bm{k}} + \Hc)$.
Here, $F(t)$ is the weak light field irradiated from the outside and $d_{\text{cv}}$ is the dipole matrix element.
Within the linear response \cite{Shimizu10}, $G(\omega)$ is estimated with the ladder approximation \cite{Haug84}.
Figure~\ref{fig:fig4} (a)-(f) shows the gain spectra corresponding to Fig.~\ref{fig:fig3} (a)-(f), respectively.
In the case of $\hbar\kappa = 0.1~{\rm\mu eV}$, two absorption peaks can be found, which result from the two flattened dispersions shown in Fig.~\ref{fig:fig1} (a).
Therefore, the separation of the peaks corresponds to $\min[4\hbar\Ek]$, the sum of the gaps in the conduction and valence bands.
Here, we note that absorption dominates the spectra because there is no or little population inversion ($n_{\text{e},k} > 0.5$) for the condensed phases in equilibrium (Fig.~\ref{fig:fig3} (a)-(c)).
In the case of $\hbar\kappa = 100~{\rm\mu eV}$, however, gain appears when the system enters into the lasing phase (Fig.~\ref{fig:fig4} (e) and (f)) although absorption still dominates in the quasi-equilibrium case (Fig.~\ref{fig:fig4} (d)).
The spectral hole (or gap) with a separation of $\min[4\hbar\Ek]$ in Fig.~\ref{fig:fig4} (f) reflects the gap formed in the renormalized band (Fig.~\ref{fig:fig1} (d)).
The existence of the gain after the second threshold is due to the population inversion for lasing (Fig.~\ref{fig:fig3} (e) and (f)) and,
as a result, can be used to distinguish the two types of the second threshold in experiments.

%%%%%%%%%%%%%%%%%%%%%%%%%%%%%%%%%%%%%%%%%%%%%%%%%%%%%%%%%%%%%%%%%%%%%%%%%%%%%%%%%%%%%%%%%%%%%%%%%%%%%%

To summarize, we have shown that there are two different types for the second threshold.
In both cases, dissociations of bound e-h pairs do not occur due to the light-induced pairing, in contrast to earlier expectations.
The gain spectra are also studied and the existence of the gain would be useful to distinguish the two different types of the second threshold.
\begin{acknowledgments}
We are grateful to T. Horikiri, Y. Shikano, M. Bamba, T. Yuge, K. Asano, T. Ohashi, H. Akiyama, M. Kuwata-Gonokami, T. Inagaki, P. Huai, H. Ajiki, J. Keeling and P. B. Littlewood for fruitful discussions.
This work is supported by the JSPS through its FIRST Program, and DYCE, KAKENHI 20104008.
\end{acknowledgments}

%%%%%%%%%%%%%%%%%%%%%%%%%%%%%%%%%%%%%%%%%%%% References and Notes %%%%%%%%%%%%%%%%%%%%%%%%%%%%%%%%%%%%%%%%%%%%%%%%%%%%

\begin{widetext}
\end{widetext}

\newpage
%%%%%%%%%%%%%%%%%%%%%%%%%%%%%%%%%%%%%%%%%%%% Supplementary material %%%%%%%%%%%%%%%%%%%%%%%%%%%%%%%%%%%%%%%%%%%%%%%%%%%%
\appendix

\section{\begin{flushleft}Supplemental Material for:\end{flushleft}
Second Thresholds in BEC-BCS-Laser Crossover of Exciton-Polariton Systems}

In this supplementary material, we describe a few details and the excitonic effects in our formalism.
In order to make the paper self-contained, we also show that the BCS gap equation as well as the MSBE can be recovered from our formalism.
Finally, we describe the parameters used in the numerical calculations and compare the results with our own previous study [Yamaguchi {\it et al}, New J. Phys. {\bf 14}, 065001 (2012)].

\subsection{A few details of the formalism}
In the definition of $p_{\bm{k}}^0$ in the main text, we have used the retarded Green's functions $G_{\alpha\alpha',\bm{k}}^{\text{R}}(\nu)$.
By using the model Hamiltonians (Eqs.~\eqref{eq:oHS}-\eqref{eq:oHSR}), the following expression of $G_{\alpha\alpha',\bm{k}}^{\text{R}}(\nu)$ can be easily obtained within the Hartree-Fock approximation (HFA):
\begin{align}%% Equations %%
G^{\text{R}}_{\bm{k}}(\nu)=
\begin{pmatrix}
\hbar\nu - \hbar\tilde{\xi}_{\text{e},\bm{k}} + \ii\hbar\gamma & \hbar\Dk\\
\hbar\Dk^{*} & \hbar\nu + \hbar\tilde{\xi}_{\text{h},\bm{k}} + \ii\hbar\gamma
\end{pmatrix}^{-1},
\label{eq:GR}%% Equations %%
\end{align}
where $G^{\text{R}}_{\bm{k}}(\nu)$ takes the matrix form of
\begin{align}%% Equations %%
G^{\text{R}}_{\bm{k}}(\nu) \equiv
\begin{pmatrix}
G^{\text{R}}_{\text{cc},\bm{k}}(\nu) & G^{\text{R}}_{\text{cv},\bm{k}}(\nu)\\
G^{\text{R}}_{\text{vc},\bm{k}}(\nu) & G^{\text{R}}_{\text{vv},\bm{k}}(\nu)
\end{pmatrix}.
\label{eq:GR_def}%% Equations %%
\end{align}
In the derivation, we have used the following relation between $\gamma$ and $\varGamma_{\bm{k}}^{\alpha}$ in Eq.~\eqref{eq:oHSR}:
\begin{align}%% Equations %%
\gamma &\cong \gamma_{\alpha,\bm{k}} \cong \pi|\varGamma_{\bm{k}}^{\alpha}|^2D_{\alpha}^{\text{B}}(\omega ),
\label{eq:gamma}
\end{align}
with the density of states defined as
\begin{align}%% Equations %%
D_{\alpha}^{\text{B}}(\omega ) &\equiv \sum_{\bm{k}}\delta(\omega_{\alpha,\bm{k}}^{\text{B}}-\omega),
\label{eq:DOScv}
\end{align}
where the dependence on the wavenumber is neglected in Eq.~\eqref{eq:gamma} for simplicity \cite{Yamaguchi12_S}.
We note that the photon loss rate $\kappa$ in Eq.~\eqref{eq:a0} has a similar relation with $\zeta_{\bm{q}}$ in Eq.~\eqref{eq:oHSR}
\begin{align}%% Equations %%
\kappa &\cong \kappa_{\bm{q}} \cong \pi|\zeta_{\bm{q}}|^2D_{\text{ph}}^{\text{B}}(\omega ),
\label{eq:kappa}
\end{align}
where the photonic density of state is defined as
\begin{align}%% Equations %%
D_{\text{ph}}^{\text{B}}(\omega ) &\equiv \sum_{\bm{p}}\delta(\omega_{\bm{p}}^{\text{B}}-\omega).
\label{eq:DOSphoton}
\end{align}
Then, through Eq.~\eqref{eq:GR}, the single-particle spectral function $A_{\text{cc/vv}}(\nu;\bm{k})$ in Eq.~\eqref{eq:Accvv} can also be derived from the definition
\begin{align}%% Equations %%
A_{\alpha\alpha'}(\nu;\bm{k}) \equiv \ii(G^{\text{R}}_{\alpha\alpha',\bm{k}}(\nu ) - G^{\text{R}*}_{\alpha'\alpha,\bm{k}}(\nu)).
\label{eq:Adef}
\end{align}
Thus, Eqs.~\eqref{eq:a0}-\eqref{eq:nehk} with Eqs.~\eqref{eq:GR} and \eqref{eq:Adef} can achieve a closed set of equations.
Here, we note that these equations are equivalent to our previous formalism \cite{Yamaguchi12_S} even though the expressions appear quite different.
One of the advantages of the present formalism is, however, that one can readily confirm the excitonic effects included in our formalism, which is described in the next Section.

%%%%%%%%%%%%%%%%%%%%%%%%%%%%%%%%%%%%%%%%%%%%%%%%%%%%%%%%%%%%%%%%%%%%
%%%%%%%%%%% Excitonic effects in the low density limit %%%%%%%%%%%%%
%%%%%%%%%%%%%%%%%%%%%%%%%%%%%%%%%%%%%%%%%%%%%%%%%%%%%%%%%%%%%%%%%%%%
\subsection{Excitonic effects in the low density limit}
In semiconductor exciton-polariton systems, the excitonic effects play quite important roles in the formation of Coulomb-bound e-h pairs (excitons) and exciton-polaritons.
In this Section, we, therefore, confirm the excitonic effects in our formalism, in order to make the paper understandable to a wide range of readers, including experimentalists.

For this purpose, we now assume that the density of electrons and holes are sufficiently low ($n_{\text{e},\bm{k}}, n_{\text{h},\bm{k}} \ll 1$ or $N_{\bm{k}} \cong -1$) with no pumping and loss ($\gamma = 0$ and $\kappa = 0$).
Under this condition, $2\xiehk{+}$ in Eq.~\eqref{eq:pk} can be written as
\begin{align}%% Equations %%
2\hbar\xiehk{+} &= \hbar\omega_{\text{e},\bm{k}} + \hbar\omega_{\text{h},\bm{k}} - \mu = \frac{\hbar^2k^2}{2m_{\text{r}}}+E_g-\mu,
\label{eq:temp1}
\end{align}
where $1/m_{\text{r}}=1/m_{\text{e}}+1/m_{\text{h}}$.
Then, Eqs.~\eqref{eq:a0} and \eqref{eq:pk} can be described as
\begin{align}%% Equations %%
0&=-(\hbar\omega_{\text{ph},0}-\mu)a_0+ g \sum_{\bm{k}} p_{\bm{k}},
\label{eq:a0_}\\
0&=-\left( \frac{\hbar^2k^2}{2m_{\text{r}}}+E_g-\mu \right) p_{\bm{k}} + \hbar g^{*}a_0 + \sum_{\bm{k}'}U'_{\bm{k}'-\bm{k}}p_{\bm{k}'},
\label{eq:pk_}
\end{align}
where the definition of the composite order parameter 
\begin{align}%% Equations %%
\Dk \equiv g^*a_0 + \hbar^{-1}\sum_{\bm{k}'}U'_{\bm{k}'-\bm{k}}p_{\bm{k}'},
\label{eq:Dk}
\end{align}
is used.
Especially, for $g = 0$ in Eq.~\eqref{eq:pk_}, we obtain
\begin{align}%% Equations %%
\frac{\hbar^2k^2}{2m_{\text{r}}} p_{\bm{k}} - \sum_{\bm{k}'}U'_{\bm{k}'-\bm{k}}p_{\bm{k}'}=-(E_g-\mu) p_{\bm{k}},
\label{eq:Wannier}
\end{align}
which is nothing but the Schr$\ddot{\text{o}}$dinger equation in $\bm{k}$-space for the single exciton bound state \cite{Yamaguchi12_S,Comte82_S,Haug-Koch_S}.
This means that the Coulomb-bound e-h pairs (excitons) can be formed in the low density limit in our formalism.
In such a case, $p_{\bm{k}}$ can be described by the bound state e-h pair wave-function $\phi_{\bm{k}}$ ($p_{\bm{k}} = \eta \phi_{\bm{k}}$ with $\sum_{\bm{k}}|\phi_{\bm{k}}|^2 = 1$) with $\mu = \hbar \omega_{\text{ex}}$, where $\hbar \omega_{\text{ex}}$ is the energy level of the exciton and the binding energy corresponds to $\Eg - \hbar \omega_{\text{ex}}$.
Our theory, thus, includes the formation of excitons.

Although $p_{\bm{k}}$ is changed from the exciton wave-function $\phi_{\bm{k}}$ by the photon-mediated attraction in the case of $g \neq 0$, it is instructive to consider the case where such an effect is not so large.
In this limit, by substituting $p_{\bm{k}} \cong \eta\phi_{\bm{k}}$ into Eqs.~\eqref{eq:a0_} and \eqref{eq:pk_}, we obtain 
\begin{align}%% Equations %%
0&=(\mu - \hbar\omega_{\text{ph},0})a_0 + \hbar g_{\text{ex}}\eta, \\
0&=(\mu - \hbar\omega_{\text{ex}})\eta + \hbar g_{\text{ex}}^{*}a_0, 
\label{eq:eigen}
\end{align}
where $g_{\text{ex}} \equiv g\sum_{\bm{k}}\phi_{\bm{k}}=g\phi_{\text{ex}}(\bm{r}=0)$ is the coupling constant renormalized by the exciton wave-function.
Then, $\mu$ is given by one of the eigenvalues of these two coupled equations, which are the eigen-frequencies of the upper and lower polaritons:
\begin{align}%% Equations %%
\omega_{\text{UP/LP}} = \frac{\omega_{\text{ph},0}+\omega_{\text{ex}} \pm \sqrt{(\omega_{\text{ph},0}-\omega_{\text{ex}})^2+4|g_{\text{ex}}|^2}}{2}. 
\label{eq:UPLP}
\end{align}
Here, $\omega_{\text{UP}}$ and $\omega_{\text{LP}}$ in Eq.~\eqref{eq:UPLP} are the well-known expressions obtained when the excitons are treated as simple bosons \cite{Haug-Koch_S}.
This means that the formation of exciton-polaritons are also included in the theory.
The excitonic effects are, thus, taken into account in our formalism within the HFA.
We note that the procedure shown here is basically the same as Section 2.1.2 in Ref.~\cite{Yamaguchi12_S}.
It is instructive to note that $\mu$ obtained by the thermal-equilibrium theories \cite{Kamide10_S,Byrnes10_S} approaches $\hbar\omega_{\text{LP}}$ in the low density limit (the black line in Fig.~\ref{fig:fig2} (b)) because the chemical potential corresponds to the lowest energy level of exciton-polaritons in the BEC phase.

%%%%%%%%%%%%%%%%%%%%%%%%%%%%%%%%%%%%%%%%%%%%%%%%%%%%%%%%%%%%%%%%%%%%
%%%%%%%%%%% Quasi-equilibrium v.s. non-equilibrium %%%%%%%%%%%%%%%%%
%%%%%%%%%%%%%%%%%%%%%%%%%%%%%%%%%%%%%%%%%%%%%%%%%%%%%%%%%%%%%%%%%%%%
\subsection{Quasi-equilibrium v.s. non-equilibrium}
In the main text, the following condition (I) is used to distinguish whether the system is in quasi-equilibrium or not:
\begin{align*}
\text{(I)}& \;\;  \min[2\hbar\Ek] \gtrsim \muB - \mu + 2\hbar\gamma + 2\kB T.
\end{align*}
Here, the system is in quasi-equilibrium if this condition (I) is satisfied.
In contrast, the system is in non-equilibrium if this condition is not satisfied.
Moreover, in such a non-equilibrium situation, it is noted that there appear $\bm{k}$-regions described by MSBE (Maxwell-Semiconductor-Bloch equation) if the following condition (II) is satisfied:
\begin{align*}
\text{(II)}& \;\; \muB-\mu \gtrsim \min[2\hbar\Ek] + 2\hbar\gamma + 2\kB T.
\end{align*}

In this paper, thus, the subject of interest ranges from quasi-equilibrium to non-equilibrium, and therefore, it is important to correctly recognize the situations under study.
To this end, in this Section, we describe the concepts and the physical meanings of the above conditions for clarity, based on the presented formalism (Eqs.~\eqref{eq:a0}-\eqref{eq:nehk} with Eqs.~\eqref{eq:GR} and \eqref{eq:Adef}).
Although discussions presented here partly overlap with our previous results~\cite{Yamaguchi12_S}, these discussions would be helpful to readers in understanding the formalism.
After that, we describe several remarks on our formalism and terminology.
In the following, $\omega_{\text{e},\bm{k}}=\omega_{\text{h},\bm{k}}$ and a charge neutrality $\mueB = \muhB$ are assumed for simplicity.

%%%%%%%%%%%%%%%%%%%%%%%%%%%%%%%%%%%%%%%%%%%%%%%%%%%%%%%%%%%%%%%%%%%%
%%%%%%%%%%%%%% Recovery of the BCS gap equation %%%%%%%%%%%%%%%%%%%%
%%%%%%%%%%%%%%%%%%%%%%%%%%%%%%%%%%%%%%%%%%%%%%%%%%%%%%%%%%%%%%%%%%%%
\subsubsection{Recovery of the BCS gap equation}
First, the condition (I) is discussed here.
By focusing on the form of $p_{\bm{k}}^0$ and $n_{\text{e/h},\bm{k}}^0$ in the main text, one can notice that these values are determined by the integrals of the bath Fermi distributions multiplied by $G_{\alpha\alpha',\bm{k}}^{\text{R}}(\nu)$ or $A_{\alpha\alpha'}(\nu;\bm{k})$.
From Eqs.~\eqref{eq:GR} and \eqref{eq:Adef}, it is easy to confirm that $G_{\alpha\alpha',\bm{k}}^{\text{R}}(\nu)$ and $A_{\alpha\alpha'}(\nu;\bm{k})$ have poles around $\nu = \pm\Ek$.
Therefore, if the condition (I) is satisfied, the bath Fermi distributions vary slowly compared with $\gamma$ around the poles (see also Fig.~\ref{fig:fig1} (c) in the main text).
This means that $f_{\text{e/h}}^{\text{B}}(\nu)$ can be approximated by the values at $\nu = \pm\Ek$ in the integrals of $p_{\bm{k}}^0$ and $n_{\text{e/h},\bm{k}}^0$.
Then, after some calculations, one can obtain
\begin{align}%% Equations %%
p_{\bm{k}}^0 &\cong \frac{\Dk}{2\Ek} \tanh \left( \frac{\beta\hbar\Ek}{2} \right),
\label{eq:temp1}\\
n_{\text{e},\bm{k}}^0&=n_{\text{h},\bm{k}}^0 \cong \frac{1}{2}-\frac{\xiehk{+}}{2\Ek} \tanh \left( \frac{\beta\hbar\Ek}{2} \right),
\label{eq:temp2}
\end{align}
where $f_{\alpha}^{\text{B}}(\pm\Ek)$ ($\alpha \in \{\text{e, h}\}$) is approximated as
\begin{align}%% Equations %%
f_{\alpha}^{\text{B}}(\pm\Ek) \cong \frac{1}{\exp (\pm\beta\hbar\Ek)+1},
\label{eq:temp3}
\end{align}
by considering the condition (I).
Then, substitutions of Eqs.~\eqref{eq:temp1} and \eqref{eq:temp2} into Eqs.~\eqref{eq:pk} and \eqref{eq:nehk} yield the following expressions for $p_{\bm{k}}$ and $n_{\text{e/h},\bm{k}}$:
\begin{align}%% Equations %%
p_{\bm{k}} &= \frac{\Dk}{2\Ek} \tanh \left( \frac{\beta\hbar\Ek}{2} \right),
\label{eq:pk_equil}\\
n_{\text{e},\bm{k}}&=n_{\text{h},\bm{k}} = \frac{1}{2}-\frac{\xiehk{+}}{2\Ek} \tanh \left( \frac{\beta\hbar\Ek}{2} \right).
\label{eq:nehk_equil}
\end{align}
In the form of Eqs.~\eqref{eq:pk_equil} and \eqref{eq:nehk_equil}, $\gamma$ does not appear even though $\gamma$ is included in Eqs.~\eqref{eq:pk} and \eqref{eq:nehk}.
This is because $\gamma$ has been canceled down by assuming $\gamma \ne 0$.
In the case of $\gamma = 0$, however, Eq.~\eqref{eq:pk_equil} and \eqref{eq:nehk_equil} cannot be obtained.
As discussed later, this is related to thermalization of the system.
Here, by using the definition of $\Dk$ (Eq.~\eqref{eq:Dk}), Eq.~\eqref{eq:a0} and Eq.~\eqref{eq:pk_equil} can be combined into one equation:
\begin{align}%% Equations %%
\hbar\Dk = \sum_{\bm{k}'}U_{ \bm{k}',\bm{k}}^{\text{eff},\kappa} \frac{\varDelta_{\bm{k}'}}{2E_{\bm{k}'}} \tanh \left( \frac{\beta\hbar E_{\bm{k}'}}{2} \right),
\label{eq:BCS}%
\end{align}
with
\begin{align}%% Equations %%
U_{ \bm{k}',\bm{k}}^{\text{eff},\kappa} \equiv \frac{\hbar |g|^2}{\xi_{\text{ph},0}-\ii\kappa} + U'_{\bm{k}'-\bm{k}}.
\label{eq:Ueff}
\end{align}
Eq.~\eqref{eq:BCS} is nothing but the gap equation in the BCS theory with the effective e-h attractive potential $U_{ \bm{k}',\bm{k}}^{\text{eff},\kappa}$.
In this sense, $\beta = 1/k_{\text{B}}T$ and $\mu$ can be regarded as the inverse temperature and the chemical potential of the {\it system}, respectively, even though $\beta$ and $\mu$ are originally introduced as the inverse temperature of the pumping baths and the oscillation frequency of the photon and polarization fields.
This means that the system can be regarded as quasi-equilibrium because thermodynamic variables of the system can be defined.
Hence, we refer to this regime as a quasi-equilibrium regime (red plots in Fig.~\ref{fig:fig2}).
In particular, for the grand state ($T \to 0$, $\kappa \to 0$), the BCS gap equation (Eq.~\eqref{eq:BCS}) with the number equation (Eq.~\eqref{eq:nehk_equil}) can recover the BEC-BCS crossover theories of the exciton-polariton condensates~\cite{Kamide10_S,Byrnes10_S}.
As a result, the calculated distributions $n_{\text{e},k}$ and polarizations $|p_{k}|$ in Fig.~\ref{fig:fig3} (a)-(d) coincide with the equilibrium theories.
The photonic polariton BEC is one of such quasi-equilibrium phases achieved by the photon-mediated e-h attraction~\cite{Note1_S}.
We emphasize, however, that the BCS gap equation (Eq.~\eqref{eq:BCS}) cannot be obtained in the case of $\gamma = 0$ because $p_{\bm{k}}$ and $n_{\text{e/h},\bm{k}}$ are not given by Eq.~\eqref{eq:pk_equil} and \eqref{eq:nehk_equil}, as mentioned above.
This means that it is essential to take the thermalization process into account for the recovery of the BCS gap equation.

Now, the physical meaning of the condition (I) can be discussed.
Since $\mu$ is equivalent to the chemical potential of the system under the present condition, $\mu_B - \mu \neq 0$ corresponds to the chemical non-equilibrium between the system and the pumping baths even if the system is in quasi-equilibrium.
In other words, $\mu_B - \mu \neq 0$ means that there is continuous flux of particles.
By considering that $\min[2\hbar\Ek]$ is the minimum energy required for breaking e-h pairs, the physical meaning of the condition (I) becomes clear; this is a condition that the particle flux ($=\mu_B - \mu$), thermalization-induced dephasing ($=2\hbar\gamma$), and temperature effect ($=2k_{\text{B}}T$), do not contribute to the dissociation of the e-h pairs. 

%%%%%%%%%%%%%%%%%%%%%%%%%%%%%%%%%%%%%%%%%%%%%%%%%%%%%%%%%%%%%%%%%%%%
%%%%%%%%%%%%%%%%%%%% Recovery of the MSBE %%%%%%%%%%%%%%%%%%%%%%%%%%
%%%%%%%%%%%%%%%%%%%%%%%%%%%%%%%%%%%%%%%%%%%%%%%%%%%%%%%%%%%%%%%%%%%%
\subsubsection{Recovery of the MSBE}
Next, the condition (II) is discussed.
For this purpose, we now consider $\bm{k}$-regions which satisfies a slightly different condition
\begin{align*}
\text{(II')}& \;\; \muB-\mu \gtrsim 2\hbar\Ek + 2\hbar\gamma + 2\kB T,
\end{align*}
because there are such $\bm{k}$-regions whenever the condition (II) is fulfilled.
In these $\bm{k}$-regions, $f_{\text{e/h}}^{\text{B}}(\pm\nu)$ varies slowly compared with $\gamma$ for $-\hbar\Ek < \hbar\nu < \hbar\Ek$ (see also Fig.~\ref{fig:fig1}(d) in the main text).
Therefore, in the integrals of $p_{\bm{k}}^0$ and $n_{\text{e/h},\bm{k}}^0$, it is possible to approximate $f_{\alpha}^{\text{B}}(\pm\nu) \cong f_{\alpha}^{\text{B}}(\xiehk{+})$ ($\alpha \in \{ \text{e, h} \}$).
As a result, we can obtain $p_{\bm{k}}^0 \cong 0$ and $n_{\text{e/h},\bm{k}}^0 \cong f_{\text{e/h},\bm{k}}$ with
\begin{align}%% Equations %%
f_{\text{e/h},\bm{k}} \equiv \frac{1}{\exp ( \beta ( \hbar\omega_{\text{e/h},\bm{k}} + \hbar\varSigma_{\text{e/h},\bm{k}}^{\text{BGR}} - \mu_{\text{e/h}}^{\text{B}} ) + 1 ) }.
\label{eq:temp5}
\end{align}
Then, Eqs.~\eqref{eq:a0}-\eqref{eq:nehk} can be described as
\begin{align}%% Equations %%
&0=-\ii\xi_{\text{ph},0}a_0+\ii g \textstyle{\sum_{\bm{k}}} p_{\bm{k}}-\kappa a_0,
\label{eq:a0_MSBE}\\
&0=-2\ii\xiehk{+}p_{\bm{k}}-\ii\Dk N_{\bm{k}}-2\gamma p_{\bm{k}},
\label{eq:pk_MSBE}\\
&0=-2\Im[\Dk p_{\bm{k}}^{*}]-2\gamma(n_{\text{e/h},\bm{k}}-f_{\text{e/h},\bm{k}}).
\label{eq:nehk_MSBE}
\end{align}
These equations are the very MSBE under the relaxation time approximation (RTA), which describes the semiconductor laser physics, in general.
Hence, we refer to this regime as a lasing regime (blue plots in Fig.~\ref{fig:fig2}).
We note that, in this regime, thermodynamic variables of the system cannot be defined and the system is in non-equilibrium.
As a result, $\mu$ does not denote the chemical potential but the laser frequency in this regime.
Furthermore, the kinetic hole burning in Fig.~\ref{fig:fig3} (e) and (f) is one of the most characteristic signatures of non-equilibrium because this phenomenon indicates that the thermalization process can no longer supply the lost carriers at sufficient speed. 

Here, the condition (II') imply that the behavior of the system is governed by (i) a large degree of non-equilibrium due to the photon leakage and (ii) a strong excitation, for the following two reasons;
firstly, a large value of $\mu_B - \mu$ means that the system is significantly affected by the photon leakage because the system becomes chemically equilibrium with the pumping baths if there is no photon leakage, as discussed above.
Therefore, a large value of $\mu_B - \mu$ suggests that a large degree of non-equilibrium is achieved.
Secondly, it is easily confirmed that a complete population inversion of the pumping baths ($f_{\text{e},\bm{k}} + f_{\text{h},\bm{k}} - 1 \cong 1$) is always achieved for the $\bm{k}$-regions that satisfies the condition (II').
This implies that the system is strongly excited and in a high-density regime.
Thus, it is quite natural that the MSBE is reproduced in the $\bm{k}$-regions under the condition (II').

However, we stress that the condition (II') depends on the wavenumber $\bm{k}$.
As a result, there remain $\bm{k}$-regions still described by the BCS gap equation even if the condition (II) is satisfied.
The MSBE and the BCS gap equation are, thus, coupled with each other in this regime.
Hence, the obtained lasing in this regime can be referred to as the BCS-coupled lasing in a strict sense~\cite{Yamaguchi12_S}.

%%%%%%%%%%%%%%%%%%%%%%%%%%%%%%%%%%%%%%%%%%%%%%%%%%%%%%%%%%%%%%%%%%%%
%%%%%%%% Several remarks on our formalism and terminology %%%%%%%%%%
%%%%%%%%%%%%%%%%%%%%%%%%%%%%%%%%%%%%%%%%%%%%%%%%%%%%%%%%%%%%%%%%%%%%
\subsubsection{Several remarks on our formalism and terminology}
We have, thus, shown that the BCS gap equation and the MSBE can be recovered from our formalism.
Finally, in this Section, we describe several remarks on our formalism and terminology.

The formalism in this paper are basically derived by the non-equilibrium Green's function (NEGF) technique~\cite{Rammer_S}, which allows us to describe non-equilibrium as well as quasi-equilibrium phenomena.
However, for the description of the BEC-BCS-Laser crossover, it is essential to take into account the thermalization process in order to drive the system toward (quasi-) equilibrium, as discussed in Section~`Recovery of the BCS gap equation'.
In other words, any formalism cannot describe a crossover from equilibrium to non-equilibrium without such a thermalization process even if the formalism is developed through the NEGF technique.
In this sense, taking the sophisticated work by Kremp {\it et al}.~\cite{Kremp08_S} as an example, such a thermalization process has to be included in their formalism when a crossover from equilibrium to non-equilibrium is studied.

We also have to note that our formalism treats the interacting carriers within the HFA.
As a result, the formation of excitons and exciton-polaritons can be taken into account (see also Section~`Excitonic effects in the low density limit'), which is in contrast to the non-interacting model~\cite{Szymanska06_S,Keeling10_S}.
However, it is well known that correlation effects beyond the HFA play important roles when discussing the connections to the BEC phase, in particular in high-temperature regimes around the critical temperature~\cite{Nozieres85_S,Ohashi02_S,Ohashi03_S}.
In this sense, we note that such correlation effects are left as future work.

Finally, we describe our terminology~\cite{Yamaguchi12_S} in order to clearly explain the situations under study.
In general, the laser physics cannot be described without non-equilibrium parameters because the steady state is determined by the balance of the pumping and loss.
Furthermore, thermodynamic variables of the system cannot be defined.
This is in contrast to quasi-equilibrium, and therefore, the laser physics cannot be described by (quasi-) equilibrium theories.
In the exciton-polariton community, the terms {\it laser} and {\it lasing} are occasionally used even for a condensation dominated by thermodynamics if the interest is in fabricating a device~\cite{Kasprzak08_S} because the system is not in true thermal equilibrium due to the pumping and loss.
However, in this paper, these terms are used only when the condensation is inherently governed by non-equilibrium kinetics, in accordance with Ref.~\cite{Bajoni08_S}.

As described in Section~`Recovery of the BCS gap equation', the system is in quasi-equilibrium under the condition (I) even if it is in chemical non-equilibrium with the pumping baths.
This regime is, therefore, not the lasing (red plots in Fig.~\ref{fig:fig2}), based on the above discussion.
In contrast, the system is in non-equilibrium if the condition (I) is not satisfied.
In such a non-equilibrium regime, MSBE (kinetic equation) plays an important role if the condition (II) is achieved, as described in Section~`Recovery of the MSBE'.
Lasing is, thus, achieved in this regime (blue plots in Fig.~\ref{fig:fig2}).
We note, however, that there is a regime which does not satisfy either of the conditions (I) and (II).
In this regime, the system is in non-equilibrium but there is no $\bm{k}$-region described by the MSBE.
Such a regime is plotted by green in Fig.~\ref{fig:fig2} and might be called a crossover regime.

However, we have to note that these criteria are based on the single-time static behaviors of the system.
Non-equilibrium signatures can still be observed in its dynamic behaviors and/or two-time correlation functions~\cite{Roumpos12_S,Belykh13_S} even though the system is in the quasi-equilibrium regime.

%%%%%%%%%%%%%%%%%%%%%%%%%%%%%%%%%%%%%%%%%%%%%%%%%%%%%%%%%%%%%%%%%%%%
%%%%%%%%%%% Parameters in the numerical calculations %%%%%%%%%%%%%%%
%%%%%%%%%%%%%%%%%%%%%%%%%%%%%%%%%%%%%%%%%%%%%%%%%%%%%%%%%%%%%%%%%%%%
\subsection{Parameters in the numerical calculations}
In our numerical calculations, the $\bm{k}$-dependence of $\Dk$ is eliminated by using a contact potential $\U{q} = U$ and the other parameters are $\hbar\omega_{\text{e},\bm{k}} = \hbar\omega_{\text{h},\bm{k}} = \hbar^2k^2/2m + \Eg/2 $, $m = 0.068m_0$ ($m_0$ is the free electron mass), $\mueB = \muhB$, $T$ = 10 K, and $\hbar\gamma$ = 4 meV.
In this context, we note that the calculations are qualitative even though these parameters are taken as realistic as possible.
Here, the contact potential and the coupling constant are, respectively, set as $U = 2.66 \times 10^{-10}$ eV and $\hbar g = 6.29 \times 10^{-7}$ eV with cut-off wavenumber $k_{\text{c}} = 1.36 \times 10^{9}$ $m^{-1}$.
In this case, the (1S) exciton level ($= \hbar\omega_{\text{ex}}$) is formed at 10 meV below $\Eg$ ($\hbar\omega_{\text{ex}} = \Eg - 10$ meV) and the lower polariton level $\hbar\omega_{\text{LP}}$ is created at 20 meV below $\Eg$ ($\hbar\omega_{\text{LP}} = \Eg - 20$ meV) under the resonant condition $\hbar\omega_{\text{ph},0} = \hbar\omega_{\text{ex}}$.
Although, in Section~`Excitonic effects in the low density limit', we have shown that $\omega_{\text{UP/LP}}$ is given by Eq.~\eqref{eq:UPLP} in the negligible limit of the photon-mediated attraction, we have determined the value of $\omega_{\text{LP}}$ with taking into account the photon-mediated effects in the numerical calculations. 

%%%%%%%%%%%%%%%%%%%%%%%%%%%%%%%%%%%%%%%%%%%%%%%%%%%%%%%%%%%%%%%%%%%%
%%%%%%%%%%%%% Comparisons with our previous results %%%%%%%%%%%%%%%%
%%%%%%%%%%%%%%%%%%%%%%%%%%%%%%%%%%%%%%%%%%%%%%%%%%%%%%%%%%%%%%%%%%%%
\subsection{Comparisons with our previous results}
In the present study, there are mainly two differences from the previous conditions of numerical calculations reported in Ref.~\cite{Yamaguchi12_S}.
The first difference is the cavity detuning; the cavity resonance $\hbar\omega_{\text{ph},0}$ is tuned to the exciton level $\hbar\omega_{\text{ex}}$ in the present case ($\hbar\omega_{\text{ph},0} = \hbar\omega_{\text{ex}} = \Eg - 10$ meV), whereas it is largely detuned from the exciton level in Ref.~\cite{Yamaguchi12_S} ($\hbar\omega_{\text{ph},0} = \hbar\omega_{\text{ex}} + 40$ meV $= \Eg + 30$ meV).
The second difference is the value of the coupling constant $\hbar g$ which determines the position of the lower polariton level under the resonant condition; the lower polariton level is located at 10 meV below the exciton level in the present case ($\hbar\omega_{\text{LP}} = \hbar\omega_{\text{ex}} - 10$ meV), while it is at 5 meV below the exciton level in Ref.~\cite{Yamaguchi12_S} ($\hbar\omega_{\text{LP}} = \hbar\omega_{\text{ex}} - 5$ meV).
Based on these differences~\cite{Note2_S}, we compare the previous and present numerical results here.

In the previous case, as a result of the large detuning and the small coupling constant, the cavity has little influence on the behavior of excitons formed in the low density regime.
Only in a high density regime, the cavity has large influence on the behavior of the condensation but excitons no longer exist in the high density regime.
Such signatures can be found in Fig.~10 and Fig.~11 of Ref.~\cite{Yamaguchi12_S}.
Red and blue plots in Fig.~10 (a) show that the order parameter $\hbar\varDelta$ comes up around $\muB - \Eg = 10$ meV when $\muB$ is increased, as indicated by the arrow (A).
This means that the condensation starts when $\muB$ reaches around the exciton level, i.e. $\muB = \hbar\omega_{\text{ex}} = \Eg - 10$ meV and there is, indeed, little influence of the cavity.
In addition, Fig.~11 (A) shows that $n_{\text{e},k}$ and $|p_{k}|$ monotonically decrease in the $k$-space.
Therefore, the condensation indicated by (A) in Fig.~10 (a) is identified as the exciton BEC.
The negligibly small photonic fraction around the point (A) in Fig.~11 is also consistent with this interpretation.
Here, we note that the photonic fraction is still small even though $\muB$ is increased up to around the point (B).
Hence, the effect of the cavity is still small at this point.
However, in contrast to Fig.~11 (A), Fig.~11 (B) shows that $n_{\text{e},k}$ has a rounded shape of the Fermi function and $|p_{k}|$ has a peak around the Fermi surface.
Therefore, the condensation around the point (B) can be classified as the e-h BCS phase.
It should be noted that, at this stage, there is no exciton anymore because the density is increased as much as the Fermi surface is formed.
This means that the cavity has little influence on the crossover from the exciton BEC to the e-h BCS phase.
In this regime, it is natural that there is little effect of the cavity loss $\kappa$ in Fig.~10 (a)-(c) (the difference between the red and blue plots for $\muB \lesssim \hbar\omega_{\text{ph},0}$).
This effect becomes apparent only after $\muB$ is increased up to around the cavity resonance ($\muB \gtrsim \hbar\omega_{\text{ph},0}$) but excitons no longer exist in this situation, as described above.

In the present case, on the other hand, the cavity has large influence on the condensation even in the low density regime due to the resonance condition and the larger coupling constant.
In Fig.~\ref{fig:fig2} (a), it can be found that the condensation occurs when $\muB$ is around $\hbar\omega_{\text{LP}}$ rather than the exciton level.
Given the monotonic decrease of $n_{\text{e},k}$ and $|p_{k}|$ in $k$-space, as shown in Fig.~\ref{fig:fig3} (a) and (d), the condensation can be identified as the exciton-polariton BEC.
As a result, the cavity loss $\kappa$ also has a great impact on the behavior of the condensation, as evidenced in Fig.~\ref{fig:fig2} (a)-(c) (the differences between the plots of $\kappa$ = 0.1 $\mu$eV and $\kappa$ = 100 $\mu$eV).
This is in contrast to the previous results where $\kappa$ plays no role in the low density regime ($\muB \lesssim \hbar\omega_{\text{ph},0}$ in Fig.~10 (a)-(c) of Ref.~\cite{Yamaguchi12_S}).
However, when $\muB$ is largely increased, the situation is not so different from the previous calculations.
This is because the band gap renormalization decreases the effective band gap $\Eg^{*}$ and the cavity level enters into the conduction and valence bands.
Therefore, for example, the behaviors of $n_{\text{e},k}$ and $|p_{k}|$ after the second thresholds in Fig.~\ref{fig:fig3} are similar to those in Fig.~11 (E) and (H) of Ref.~\cite{Yamaguchi12_S}.
Thus, from the comparison of the present and previous numerical results, we can learn that the effect of the cavity in the low density regime becomes more important in the present case than in the previous one.

%%%%%%%%%%%%%%%%%%%%%%%%%%%%%%%%%%%%%%%%%%%% Supplemental References and Notes %%%%%%%%%%%%%%%%%%%%%%%%%%%%%%%%%%%%%%%%%%%%%%%%%%%%


\begin{thebibliography}{99}
\bibitem{Note1_S}
The \textquotedblleft photonic polariton BEC\,\textquotedblright\, is originally referred to as the \textquotedblleft photonic BEC\,\textquotedblright\ in Ref.~\cite{Kamide10_S}.
However, this terminology can be easily confused with the \textquotedblleft photon BEC\,\textquotedblright\ in Ref.~\cite{Klaers10_S}.
We have, therefore, used the term of \textquotedblleft photonic polariton BEC\,\textquotedblright, instead of \textquotedblleft photonic BEC\,\textquotedblright, in order to avoid misunderstanding.
In the photonic polariton BEC, the electron-hole-photon system is in (quasi-) equilibrium.
However, in the \textquotedblleft photon BEC\,\textquotedblright, only the photon system is in (quasi-) equilibrium.
These phases are not equivalent.
\bibitem{Note2_S}
$T$ = 10 K is also different from the previous condition $T$ = 0 K but this difference does not play any significant role because $k_{\text{B}}T$ is smaller than $\hbar\gamma$.
\end{thebibliography}
\end{document}